\documentclass[twocolumn,showpacs,prc,floatfix]{revtex4}
\usepackage{graphicx}
\usepackage{amsmath}
\newcommand{\p}[1]{(\ref{#1})}
\newcommand\beq{\begin{eqnarray}} \newcommand\eeq{\end{eqnarray}}
\newcommand\beqstar{\begin{eqnarray*}} \newcommand\eeqstar{\end{eqnarray*}}
\newcommand{\beqe}{\begin{equation}} \newcommand{\eeqe}{\end{equation}}
\newcommand{\beqestar}{\begin{equation*}} \newcommand{\eeqestar}{\end{equation*}}
\newcommand{\bal}{\begin{align}}
\newcommand{\tbf}{\textbf}

\begin {document}
\title{Spin polarized states \protect\\  in
strongly asymmetric nuclear matter }
\author{ A. A. Isayev}
\affiliation{Kharkov Institute of Physics and Technology,
Academicheskaya Str. 1,
 Kharkov, 61108, Ukraine}
 \author{J. Yang}
 \affiliation{Dept. of Physics and Center for Space Science and Technology,\\
Ewha Womans University, Seoul 120-750, Korea\\ and \\
Center for High Energy Physics, Kyungbook National University, Daegu 702-701, Korea}
 \date{\today}
\begin{abstract}  The possibility of appearance of spin polarized states
in strongly asymmetric nuclear matter is analyzed within the
framework of a Fermi liquid theory with the Skyrme effective
interaction. The zero temperature dependence of the neutron and
proton spin polarization parameters as functions of density is
found for SLy4 and SLy5 effective forces. It is shown that at some
critical density strongly asymmetric nuclear matter  undergoes a
phase transition to the state with the oppositely directed spins
of neutrons and protons while the state with the same direction of
spins does not appear. In comparison with neutron matter, even
small admixture of protons strongly decreases the threshold
density of spin instability. It is clarified that protons become
totally polarized within very narrow density domain while  the
density profile of the neutron spin polarization parameter is
characterized by the appearance of long tails near the transition
density.
\end{abstract}
 \maketitle

\section{Introduction}
 The spontaneous appearance of  spin polarized states in nuclear
matter is the topic of a great current interest due to relevance
in astrophysics. In particular, the effects of spin correlations
in the medium strongly influence the neutrino cross section and
neutrino luminosity. Hence, depending on whether nuclear matter is
spin polarized or not, drastically different scenarios of
supernova explosion and cooling of neutron stars can be realized.
Another aspect relates to pulsars, which are considered to be
rapidly rotating neutron stars, surrounded by strong magnetic
field. There is still no general consensus regarding the mechanism
to generate such strong magnetic field of a neutron star. One of
the hypotheses is that magnetic field can be produced by a
spontaneous ordering of spins in the dense stellar core.

 The
possibility of a phase transition of normal nuclear matter to the
ferromagnetic state was studied by many authors. In  the gas model
of hard spheres,  neutron matter becomes ferromagnetic at
$\varrho\approx0.41\,\mbox{fm}^{-3}$~\cite{R}. It was found in
Refs.~\cite{S,O} that the inclusion of long--range attraction
significantly increases the ferromagnetic transition density
(e.g., up to $\varrho\approx2.3\,\mbox{fm}^{-3}$ in the Brueckner
theory with a simple central potential and hard core only for
singlet spin states~\cite{O}). By determining magnetic
susceptibility with Skyrme effective forces, it was shown in
Ref.~\cite{VNB} that the ferromagnetic transition occurs at
$\varrho\approx0.18$--$0.26\,\mbox{fm}^{-3}$. The Fermi liquid
criterion for the ferromagnetic instability in neutron matter with
the Skyrme interaction is reached at
$\varrho\approx2$--$4\varrho_0$~\cite{RPLP}, where
$\varrho_0=0.16\,\mbox{fm}^{-3}$ is the nuclear matter saturation
density.    The general conditions on the parameters of
neutron--neutron interaction, which  result in a magnetically
ordered state of neutron matter, were formulated in
Ref.~\cite{ALP}. Spin correlations in dense neutron matter were
studied within the relativistic Dirac--Hartree--Fock approach with
the effective nucleon--meson Lagrangian in Ref.~\cite{MNQN},
predicting the ferromagnetic transition at several times nuclear
matter saturation density. The importance of the Fock exchange
term in the relativistic mean--field approach for the occurrence
of ferromagnetism in nuclear matter was established in
Ref.~\cite{TT}. The stability of strongly asymmetric nuclear
matter with respect to spin fluctuations was investigated in
Ref.~\cite{KW}, where it was shown that  the system with localized
protons can develop a spontaneous polarization, if  the
neutron--proton spin interaction exceeds some threshold value.
This conclusion was confirmed also by calculations within the
relativistic Dirac--Hartree--Fock approach to strongly asymmetric
nuclear matter~\cite{BMNQ}. Competition between ferromagnetic (FM)
and antiferromagnetic (AFM) spin ordering in symmetric nuclear
matter with the Skyrme effective interaction was studied in
Ref.~\cite{I}, where it was clarified that FM spin state is
thermodynamically preferable over AFM one for all relevant
densities.

For the models with realistic nucleon--nucleon (NN)
interaction, the ferromagnetic phase transition seems to be
suppressed up to densities well above
$\varrho_0$~\cite{PGS}--\cite{H}. In particular, no evidence of
ferromagnetic instability has been found in recent studies of
neutron matter~\cite{VPR} and asymmetric nuclear matter~\cite{VB}
within the Brueckner--Hartree--Fock approximation with realistic
Nijmegen II, Reid93 and Nijmegen NSC97e NN interactions. The same
conclusion was obtained in Ref.~\cite{FSS}, where magnetic
susceptibility of neutron matter was calculated with the use of
the Argonne $v_{18}$ two--body potential and Urbana IX three--body
potential.

Here we  continue the study of spin polarizability of nuclear
matter with the use of an effective NN interaction. As a framework
of consideration, a Fermi liquid (FL) description of nuclear
matter is chosen~\cite{AKP,AIP}. As a potential of NN interaction,
we use the Skyrme effective interaction, utilized earlier in a
number of contexts for nuclear matter
calculations~\cite{SYK}--\cite{AAI}. The main emphasis will be
laid on strongly asymmetric nuclear matter and neutron matter as
its limiting case. We explore the possibility of FM and
 AFM phase transitions in nuclear matter, when
the spins of protons and neutrons are aligned in the same
direction or in the opposite direction, respectively. In contrast
to the approach, based on the calculation of magnetic
susceptibility, we obtain the self--consistent equations for the
FM and  AFM spin order parameters and find their solutions at zero
temperature. This allows us to determine not only the critical
density of instability with respect to spin fluctuations, but to
establish the density dependence of the order parameters and to
clarify the question of thermodynamic stability of various phases.

Note that we consider  the thermodynamic properties of spin
polarized states in nuclear matter up to the high density region
relevant for astrophysics. Nevertheless, we take into account the
nucleon degrees of freedom only,  although other degrees of
freedom, such as pions, hyperons, kaons,  or quarks could be
important at such high densities.
\section{Basic Equations}
 The normal states of nuclear matter are described
  by the normal distribution function of nucleons $f_{\kappa_1\kappa_2}=\mbox{Tr}\,\varrho
  a^+_{\kappa_2}a_{\kappa_1}$, where
$\kappa\equiv({\bf{p}},\sigma,\tau)$, ${\bf p}$ is momentum,
$\sigma(\tau)$ is the projection of spin (isospin) on the third
axis, and $\varrho$ is the density matrix of the system.  The energy
of the system is specified as a functional of the distribution
function $f$, $E=E(f)$, and determines the single particle energy
 \begin{eqnarray}
\varepsilon_{\kappa_1\kappa_2}(f)=\frac{\partial E(f)}{\partial
f_{\kappa_2\kappa_1}}. \label{1} \end{eqnarray} The
self-consistent matrix equation for determining the distribution
function $f$ follows from the minimum condition of the
thermodynamic potential \cite{AKP} and is
  \begin{eqnarray}
 f=\left\{\mbox{exp}(Y_0\varepsilon+
Y_4)+1\right\}^{-1}\equiv
\left\{\mbox{exp}(Y_0\xi)+1\right\}^{-1}.\label{2}\end{eqnarray}
Here the quantities $\varepsilon$ and $Y_4$ are matrices in the space of
$\kappa$ variables, with
$Y_{4\kappa_1\kappa_2}=Y_{4\tau_1}\delta_{\kappa_1\kappa_2}$
$(\tau_1=n,p)$, $Y_0=1/T,\ Y_{4n}=-\mu_n^0/T$ and
$Y_{4p}=-\mu_p^0/T$
 the Lagrange multipliers, $\mu_n^0$ and $\mu_p^0$  the chemical
potentials of  neutrons and protons, and $T$  the temperature.
Since it is assumed to consider a nuclear system with an excess of
neutrons, the positive isospin projection is assigned
 to neutrons. This is different from the formalism of
 Ref.~\cite{I}, aimed to investigate  symmetric nuclear matter.
 Further we shall study the possibility of
formation of various types of spin ordering (ferromagnetic and
antiferromagnetic) in nuclear matter.

The normal  distribution function can be expanded in the Pauli
matrices $\sigma_i$ and $\tau_k$ in spin and isospin
spaces
\begin{align} f({\bf p})&= f_{00}({\bf
p})\sigma_0\tau_0+f_{30}({\bf p})\sigma_3\tau_0\label{7.2}\\
&\quad + f_{03}({\bf p})\sigma_0\tau_3+f_{33}({\bf
p})\sigma_3\tau_3. \nonumber
\end{align}
 For the energy functional invariant with
respect to rotations in spin and isospin spaces, the structure of
the single particle energy  is  similar to the structure of the
distribution function $f$: \begin{align} \varepsilon({\bf p})&=
\varepsilon_{00}({\bf
p})\sigma_0\tau_0+\varepsilon_{30}({\bf p})\sigma_3\tau_0\label{7.3}\\
&\quad + \varepsilon_{03}({\bf
p})\sigma_0\tau_3+\varepsilon_{33}({\bf p})\sigma_3\tau_3.
\nonumber
\end{align}
Using Eqs.~\p{2}, \p{7.3}, one can express evidently the
distribution functions $f_{00},f_{30},f_{03},f_{33}$
 in
terms of the quantities $\varepsilon$: \begin{align}
f_{00}&=\frac{1}{4}\{n(\omega_{+,+})+n(\omega_{+,-})+n(\omega_{-,+})+n(\omega_{-,-})
\},\nonumber
 \\
f_{30}&=\frac{1}{4}\{n(\omega_{+,+})+n(\omega_{+,-})-n(\omega_{-,+})-n(\omega_{-,-})
\},\nonumber\\
f_{03}&=\frac{1}{4}\{n(\omega_{+,+})-n(\omega_{+,-})+n(\omega_{-,+})-n(\omega_{-,-})
\},\nonumber\\
f_{33}&=\frac{1}{4}\{n(\omega_{+,+})-n(\omega_{+,-})-n(\omega_{-,+})+n(\omega_{-,-})
\}.\label{2.4}
 \end{align} Here $n(\omega)=\{\exp(Y_0\omega)+1\}^{-1}$ and
\begin{gather*}
\omega_{+,+}=\xi_{00}+\xi_{30}+\xi_{03}+\xi_{33},\;\\
\omega_{+,-}=\xi_{00}+\xi_{30}-\xi_{03}-\xi_{33},\;\\
\omega_{-,+}=\xi_{00}-\xi_{30}+\xi_{03}-\xi_{33},\;\\
\omega_{-,-}=\xi_{00}-\xi_{30}-\xi_{03}+\xi_{33},\;\end{gather*}
where \begin{align*}\xi_{00}&=\varepsilon_{00}-\mu_{00}^0,\;
\xi_{30}=\varepsilon_{30},\;
\\
\xi_{03}&=\varepsilon_{03}-\mu_{03}^0,\;\xi_{33}=\varepsilon_{33},\\
\mu_{00}^0&={\frac{\mu_n^0+\mu_p^0}{2}},\quad
\mu_{03}^0={\frac{\mu_n^0-\mu_p^0}{2}}.\end{align*}
 As follows from the structure of the distribution
functions $f$, the quantity $\omega_{\pm,\pm}$, being the exponent
in the Fermi distribution function $n$, plays the role of the
quasiparticle spectrum. We consider the case when the spectrum is
four--fold split due to the spin and isospin dependence of the
single particle energy $\varepsilon({\bf p})$ in Eq.~\p{7.3}. The
branches $\omega_{\pm,+}$ correspond to neutrons with spin up and
spin down, and  the branches $\omega_{\pm,-}$  to protons with
spin up and spin down.

The distribution functions $f$ should satisfy the norma\-lization
conditions
\begin{align} \frac{4}{\cal
V}\sum_{\bf p}f_{00}({\bf p})&=\varrho,\label{3.1}\\
\frac{4}{\cal V}\sum_{\bf p}f_{03}({\bf
p})&=\varrho_n-\varrho_p\equiv\alpha\varrho,\label{3.3}\\
\frac{4}{\cal V}\sum_{\bf p}f_{30}({\bf
p})&=\varrho_\uparrow-\varrho_\downarrow\equiv\Delta\varrho_{\uparrow\uparrow},\label{3.2}\\
\frac{4}{\cal V}\sum_{\bf p}f_{33}({\bf
p})&=(\varrho_{n\uparrow}+\varrho_{p\downarrow})-
(\varrho_{n\downarrow}+\varrho_{p\uparrow})\equiv\Delta\varrho_{\uparrow\downarrow}.\label{3.4}
 \end{align}
 Here $\alpha$ is the isospin asymmetry parameter, $\varrho_{n\uparrow},\varrho_{n\downarrow}$ and
 $\varrho_{p\uparrow},\varrho_{p\downarrow}$ are the neutron and
 proton number densities with spin up and spin down,
 respectively;
 $\varrho_\uparrow=\varrho_{n\uparrow}+\varrho_{p\uparrow}$ and
$\varrho_\downarrow=\varrho_{n\downarrow}+\varrho_{p\downarrow}$
are the nucleon densities with spin up and spin down. The
quantities $\Delta\varrho_{\uparrow\uparrow}$ and
$\Delta\varrho_{\uparrow\downarrow}$ may be regarded as FM and AFM
spin order parameters. Indeed, in symmetric nuclear matter, if all
nucleon spins are aligned in one direction (totally polarized FM
spin state), then $\Delta\varrho_{\uparrow\uparrow}=\varrho$ and
$\Delta\varrho_{\uparrow\downarrow}=0$; if all neutron spins are
aligned in one direction and  all proton spins in the opposite one
(totally polarized  AFM spin state), then
$\Delta\varrho_{\uparrow\downarrow}=\varrho$ and
$\Delta\varrho_{\uparrow\uparrow}=0$. In turn, from
Eqs.~\p{3.1}--\p{3.4} one can find the
neutron and proton number densities with spin up and spin down as functions of
the total density $\varrho$, isospin excess $\delta\varrho\equiv
\alpha\varrho$, and FM and AFM order parameters
$\Delta\varrho_{\uparrow\uparrow}$ and
$\Delta\varrho_{\uparrow\downarrow}$: \bal
\varrho_{n\uparrow}&=\frac{1}{4}(\varrho+\delta\varrho+\Delta\varrho_{\uparrow\uparrow}+
\Delta\varrho_{\uparrow\downarrow}),\nonumber\\
\varrho_{n\downarrow}&=\frac{1}{4}(\varrho+\delta\varrho-\Delta\varrho_{\uparrow\uparrow}-
\Delta\varrho_{\uparrow\downarrow}),\nonumber\\
\varrho_{p\uparrow}&=\frac{1}{4}(\varrho-\delta\varrho+\Delta\varrho_{\uparrow\uparrow}-
\Delta\varrho_{\uparrow\downarrow}),\nonumber\\ 
\varrho_{p\downarrow}&=\frac{1}{4}(\varrho-\delta\varrho-\Delta\varrho_{\uparrow\uparrow}+
\Delta\varrho_{\uparrow\downarrow}).\nonumber\end{align}

In order to characterize spin ordering in the neutron and  proton
subsystems, it is convenient to introduce   neutron and proton
spin polarization parameters \beqe
\Pi_n=\frac{\varrho_{n\uparrow}-\varrho_{n\downarrow}}{\varrho_n},\quad
\Pi_p=\frac{\varrho_{p\uparrow}-\varrho_{p\downarrow}}{\varrho_p}.
\end{equation} The expressions for the spin order parameters
$\Delta\varrho_{\uparrow\uparrow}$ and
$\Delta\varrho_{\uparrow\downarrow}$ through the spin polarization
parameters read \begin{align}
\Delta\varrho_{\uparrow\uparrow}=\varrho_n\Pi_n+\varrho_p\Pi_p,\;
\Delta\varrho_{\uparrow\downarrow}=\varrho_n\Pi_n-\varrho_p\Pi_p.\nonumber
\end{align}

 To obtain the self--consistent equations, we specify the energy functional of
the system in the form
\begin{align} E(f)&=E_0(f)+E_{int}(f), \label{14}\\
{E}_0(f)&=4\sum\limits_{ \bf p}^{} \varepsilon_0({\bf
p})f_{00}({\bf p}),\;\varepsilon_0({\bf p})=\frac{{\bf
p}^{\,2}}{2m_{0}},\nonumber
\\ {E}_{int}(f)&=2\sum\limits_{ \bf p}^{}\{
\tilde\varepsilon_{00}({\bf p})f_{00}({\bf p})+
\tilde\varepsilon_{30}({\bf p})f_{30}({\bf p})\nonumber\\
&\quad+\tilde\varepsilon_{03}({\bf p})f_{03}({\bf p})+
\tilde\varepsilon_{33}({\bf p})f_{33}({\bf p})\} ,
\nonumber\end{align} \begin{align}\tilde\varepsilon_{00}({\bf
p})&=\frac{1}{2\cal V}\sum_{\bf q}U_0({\bf k})f_{00}({\bf
q}),\;{\bf k}=\frac{{\bf p}-{\bf q}}{2}, \nonumber\\
\tilde\varepsilon_{30}({\bf p})&=\frac{1}{2\cal V}\sum_{\bf
q}U_1({\bf k})f_{30}({\bf q}),\nonumber\\ 
\tilde\varepsilon_{03}({\bf p})&=\frac{1}{2\cal V}\sum_{\bf
q}U_2({\bf k})f_{03}({\bf q}), \nonumber\\
\tilde\varepsilon_{33}({\bf p})&=\frac{1}{2\cal V}\sum_{\bf
q}U_3({\bf k})f_{33}({\bf q}). \nonumber
\end{align}
 Here
  $m_0$ is the bare mass of a nucleon, $U_0({\bf k}),...,U_3({\bf k}) $ are the normal FL
amplitudes, and
$\tilde\varepsilon_{00},\tilde\varepsilon_{30},\tilde\varepsilon_{03},\tilde\varepsilon_{33}$
are the FL corrections to the free single particle spectrum.
Further we do not take into account the effective tensor forces,
which lead to coupling of the momentum and spin degrees of freedom
\cite{HJ,D,FMS}, and, correspondingly, to anisotropy in the
momentum dependence of FL amplitudes with respect to the spin
polarization axis. Using Eqs.~\p{1} and \p{14}, we
get the self--consistent equations in the form \bal \xi_{00}({\bf
p})&=\varepsilon_{0}({\bf p})+\tilde\varepsilon_{00}({\bf
p})-\mu_{00}^0,\;
\xi_{30}({\bf p})=\tilde\varepsilon_{30}({\bf p}), \\
\xi_{03}({\bf p})&=\tilde\varepsilon_{03}({\bf p})-\mu_{03}^0, \;
\xi_{33}({\bf p})=\tilde\varepsilon_{33}({\bf p}).
\nonumber\end{align}
  To obtain
 numerical results, we  use the Skyrme effective interaction.
In the case of Skyrme forces the normal FL amplitudes
read~\cite{AIP} \begin{align}
U_0({\bf k})&=6t_0+t_3\varrho^\beta
+\frac{2}{\hbar^2}[3t_1+t_2(5+4x_2)]{\bf k}^{2},
\\
U_1({\bf
k})&=-2t_0(1-2x_0)-\frac{1}{3}t_3\varrho^\beta(1-2x_3)\nonumber\\
&\quad-\frac{2}{\hbar^2}[t_1(1-2x_1)-t_2(1+2x_2) ]{\bf
k}^{2}\equiv a+b{\bf k}^{2},
\nonumber\\
U_2({\bf
k})&=-2t_0(1+2x_0)-\frac{1}{3}t_3\varrho^\beta(1+2x_3)\nonumber\\
&\quad-\frac{2}{\hbar^2}[t_1(1+2x_1)- t_2(1+2x_2)]{\bf k}^{2},\nonumber\\
U_3({\bf k})&=-2t_0-\frac{1}{3}t_3\varrho^\beta
-\frac{2}{\hbar^2}(t_1- t_2){\bf k}^{2}\equiv c+d{\bf
k}^{2},\nonumber
\end{align}
where $t_i,x_i,\beta$ are the phenomenological constants,
characterizing a given parameterization of the Skyrme forces. In
the  numerical calculations  we
  shall use   SLy4 and SLy5 potentials~\cite{CBH},
  developed to fit the properties of systems with large isospin
  asymmetry.
 With
account of the evident form of FL amplitudes and
Eqs.~\p{3.1}--\p{3.4}, one can obtain \begin{align}
\xi_{00}&=\frac{p^2}{2m_{00}}-\mu_{00}, \\
\xi_{03}&=\frac{p^2}{2m_{03}}-\mu_{03},\\
\xi_{30}&=(a+b\frac{{\bf
p}^{2}}{4})\frac{\Delta\varrho_{\uparrow\uparrow}}{8}+\frac{b}{32}\langle
{\bf q}^{2}\rangle_{30}, \label{4.3}\\
\xi_{33}&=(c+d\frac{{\bf
p}^{2}}{4})\frac{\Delta\varrho_{\uparrow\downarrow}}{8}+\frac{d}{32}\langle
{\bf q}^{2}\rangle_{33},\label{4.4}
\end{align}
where the effective nucleon mass $m_{00}$ and  effective isovector
mass
  $m_{03}$ are defined by
 the formulae:
\begin{eqnarray}
\frac{\hbar^2}{2m_{00}}&=&\frac{\hbar^2}{2m_0}+\frac{\varrho}{16}
[3t_1+t_2(5+4x_2)],\label{18}\\
\frac{\hbar^2}{2m_{03}}&=& \frac{\alpha\varrho}{16}[t_2(1+2x_2)-
t_1(1+2x_1)],\nonumber\end{eqnarray} and the renormalized chemical
potentials $\mu_{00}$ and $\mu_{03}$ should be determined from Eqs.
\p{3.1}, \p{3.3}. In Eqs. ~\p{4.3} and  \p{4.4}, $\langle {\bf
q}^{2}\rangle_{30}$ and $\langle {\bf q}^{2}\rangle_{33}$ are the second
order moments of the corresponding distribution functions
\begin{align} \langle {\bf
q}^{2}\rangle_{30}&=\frac{4}{V}\sum_{\bf q}{\bf
q}^2f_{30}({\bf q}),\label{6.1}\\
\langle {\bf q}^{2}\rangle_{33}&=\frac{4}{V}\sum_{\bf q}{\bf
q}^2f_{33}({\bf q}). \label{6.2}\end{align}

Thus, with account of the expressions \p{2.4} for the distribution
functions $f$, we obtain the self--consistent equations
\p{3.1}--\p{3.4}, \p{6.1}, and \p{6.2} for the effective chemical
potentials $\mu_{00},\mu_{03}$,
 FM  and AFM spin
 order parameters
$\Delta\varrho_{\uparrow\uparrow}$,
$\Delta\varrho_{\uparrow\downarrow}$, and  second order moments
$\langle {\bf q}^{2}\rangle_{30}, \langle {\bf q}^{2}\rangle_{33}$.
It is easy to see, that the self--consistent equations remain
invariable under a global flip of  spins, when neutrons (protons)
with spin up and spin down are interchanged, and under a global
flip of  isospins, when   neutrons and protons with the same spin
projection are interchanged.

Let us consider, what differences will be in the case of neutron
matter.  Neutron matter is an infinite nuclear system, consisting
of nucleons of one species, i.e., neutrons, and, hence, the
formalism of one--component Fermi liquid should be applied for the
description of its properties. Formally neutron matter can be
considered as the limiting case of asymmetric nuclear matter,
corresponding to the isospin asymmetry $\alpha=1$. The individual
state of a neutron is  characterized by momentum $\bf p$ and spin
projection $\sigma$. The self--consistent equation has the form of
Eq.~\p{2}, where all quantities are matrices in the space of
$\kappa\equiv(\tbf p,\sigma)$ variables. The normal distribution
function and single particle energy can be expanded in the Pauli
matrices in spin space \bal  f({\bf p})&= f_{0}({\bf
p})\sigma_0+f_{3}({\bf p})\sigma_3,
\label{7.23}\\
  \varepsilon({\bf p})&= \varepsilon_{0}({\bf p})\sigma_0+\varepsilon_{3}({\bf
p})\sigma_3. \nonumber
\end{align}

The energy functional of neutron matter is characterized by two
normal FL amplitudes $U_0^n(\tbf k)$ and $U_1^n(\tbf k)$.
 Applying the same procedure, as in Ref.~\cite{AIP}, the normal FL amplitudes
can be found in terms of the Skyrme force parameters
$t_i,x_i,\beta$:
\bal U_0^n({\bf k})&=2t_0(1-x_0)+\frac{t_3}{3}\varrho^\beta(1-x_3)\label{101}\\&\quad
+\frac{2}{\hbar^2}[t_1(1-x_1)+3t_2(1+x_2)]{\bf k}^{2},
\nonumber\\
U_1^n({\bf
k})&=-2t_0(1-x_0)-\frac{t_3}{3}\varrho^\beta(1-x_3)\label{102}\\&\quad
+\frac{2}{\hbar^2}[t_2(1+x_2)-t_1(1-x_1)]{\bf k}^{2}\equiv
a_n+b_n{\bf k}^{2}.\nonumber\end{align} With account of
Eqs.~\p{101} and \p{102}, the normalization conditions for the
distribution functions can be written in the form
\bal\frac{2}{\cal V}\sum_{\bf p}f_{0}({\bf p})&=\varrho,\;
\label{3.17}\\
\frac{2}{\cal V}\sum_{\bf p}f_{3}({\bf
p})&=\varrho_\uparrow-\varrho_\downarrow\equiv\Delta\varrho_{\uparrow\uparrow}.
\label{3.30}
\end{align}
Here  $\varrho_\uparrow$ and $\varrho_\downarrow$ are the neutron
number densities with spin up and spin down and
\begin{align}
f_{0}&=\frac{1}{2}\{n(\omega_{+})+n(\omega_{-})
\},\quad\omega_\pm=\xi_0\pm\xi_3,\label{4.38}\\
f_{3}&=\frac{1}{2}\{n(\omega_{+})-n(\omega_{-})\},\label{4.39}\\
\xi_{0}&=\frac{p^2}{2m_{n}}-\mu_{n},\\
\xi_{3}&=(a_n+b_n\frac{{\bf
p}^{2}}{4})\frac{\Delta\varrho_{\uparrow\uparrow}}{4}+\frac{b_n}{16}\langle
{\bf q}^{2}\rangle_{3}. \label{4.33}
\end{align}
The effective neutron mass $m_{n}$  is defined by
 the formula
\begin{eqnarray}
\frac{\hbar^2}{2m_{n}}&=&\frac{\hbar^2}{2m_0}+\frac{\varrho}{8}
[t_1(1-x_1)+3t_2(1+x_2)],\label{181}\end{eqnarray} and the
quantity $\langle {\bf q}^{2}\rangle_{3}$ in Eq.~\p{4.33}  is the
second order moment of the distribution function $f_3$:
\begin{align} \langle {\bf
q}^{2}\rangle_{3}&=\frac{2}{V}\sum_{\bf q}{\bf q}^2f_{3}({\bf
q}).\label{6.11}\end{align}  Thus, with account of the expressions
\p{4.38} and  \p{4.39} for the distribution functions $f$, we obtain
the self--consistent equations \p{3.17}, \p{3.30}, and  \p{6.11} for
the effective chemical potential $\mu_{n}$,
  spin  order parameter
$\Delta\varrho_{\uparrow\uparrow}$,
 and  second order moment
$\langle {\bf q}^{2}\rangle_{3}$. 
\section{Phase transitions in strongly asymmetric nuclear matter}
Early researches on spin polarizability  of nuclear matter with
the Skyrme effective interaction were based on the calculation of
magnetic susceptibility and finding  its pole
structure~\cite{VNB,RPLP}, determining the onset of   instability
with respect to spin fluctuations. Here we shall find directly
solutions of the self--consistent equations for the FM and AFM
spin order parameters as functions of density at zero temperature.
A special emphasis will be laid on the study of strongly
asymmetric nuclear matter ($\alpha\lesssim1$) while in symmetric
nuclear matter
 FM spin ordering is thermodynamically more preferable than AFM
one~\cite{I}.

If   all   neutron and proton spins are aligned in one direction,
then for nontrivial solutions of the self--consistent equations we
have \bal \Delta\varrho_{\uparrow\uparrow}&=\varrho,\quad
\Delta\varrho_{\uparrow\downarrow}=\alpha\varrho,\\
\langle {\bf q}^{2}\rangle_{30}&=\frac{3}{10}\varrho
k_F^2[(1+\alpha)^{5/3}+(1-\alpha)^{5/3}],\nonumber\\ \langle {\bf
q}^{2}\rangle_{33}&=\frac{3}{10}\varrho
k_F^2[(1+\alpha)^{5/3}-(1-\alpha)^{5/3}],\nonumber\end{align}
where
 $k_F=(3\pi^2\varrho)^{1/3}$
is the Fermi momentum  of totally polarized symmetric nuclear
matter. Therefore, for the partial number densities of nucleons
with spin up and spin down one can get
\begin{equation}
\varrho_{n\uparrow}=\frac{1+\alpha}{2}\varrho,\;
\varrho_{p\uparrow}=\frac{1-\alpha}{2}\varrho,\;
\varrho_{p\downarrow}=\varrho_{n\downarrow}=0.\end{equation}

If   all neutron spins are aligned in one direction and  all
proton spins in the opposite one, then  \bal
\Delta\varrho_{\uparrow\uparrow}&=\alpha\varrho,\quad
\Delta\varrho_{\uparrow\downarrow}=\varrho,\\
\langle {\bf q}^{2}\rangle_{30}&=\frac{3}{10}\varrho
k_F^2[(1+\alpha)^{5/3}-(1-\alpha)^{5/3}],\nonumber\\ \langle {\bf
q}^{2}\rangle_{33}&=\frac{3}{10}\varrho
k_F^2[(1+\alpha)^{5/3}+(1-\alpha)^{5/3}],\nonumber\end{align} and,
hence,
\begin{equation}
\varrho_{n\uparrow}=\frac{1+\alpha}{2}\varrho,\;
\varrho_{p\downarrow}=\frac{1-\alpha}{2}\varrho,\;
\varrho_{p\uparrow}=\varrho_{n\downarrow}=0.\end{equation}

\begin{figure}[b]
\includegraphics[height=12.6cm,width=8.6cm,trim=49mm 105mm 56mm 46mm,
draft=false,clip]{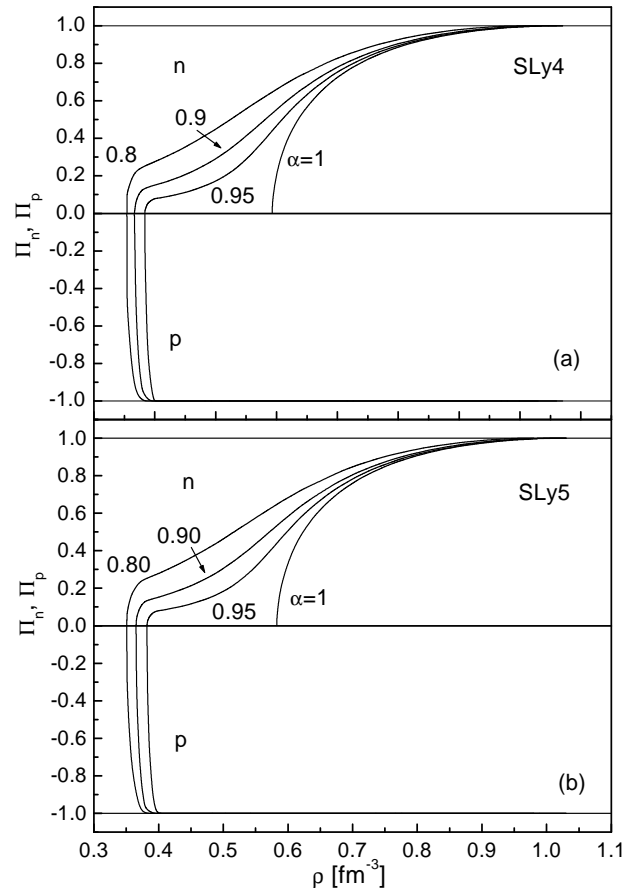} \caption{Neutron and proton spin
polarization parameters as  functions of density  at zero
temperature for (a) SLy4 force and (b) SLy5 force. }\label{fig1}
\end{figure}

Now we present the results of numerical solution of the
self--consistent equations with the effective SLy4 and SLy5 forces
for strongly asymmetric nuclear ($\alpha=0.95, 0.9, 0.8$) and
neutron ($\alpha=1$) matter.      The neutron and proton spin
polarization parameters $\Pi_n$ and $\Pi_p$  are shown in Fig.~1 as
 functions of density at zero temperature. Since in totally
polarized state the signs of spin polarizations are opposite
($\Pi_n=1,\Pi_p=-1$), considering solutions correspond to the
case, when spins of neutrons and protons are aligned in the
opposite direction. Note that for SLy4 and SLy5 forces, being
relevant for the description of strongly asymmetric nuclear
matter, there are no solutions, corresponding to the same
direction of neutron and proton spins. The reason is that the sign
of the multiplier $t_3(-1+2x_3)$ in the density dependent term of
the FL amplitude $U_1$, determining spin--spin correlations, is
positive, and, hence, corresponding term increases with the
increase of nuclear matter density, preventing instability with
respect to spin fluctuations. Contrarily, the density dependent term
$-t_3\varrho^\beta/3$ in the FL amplitude $U_3$, describing
spin--isospin correlations, is negative, leading to spin
instability with the oppositely directed spins of neutrons and
protons at higher densities.

\begin{figure}[t]
\includegraphics[height=12.6cm,width=8.6cm,trim=49mm 103mm 56mm 46mm,
draft=false,clip]{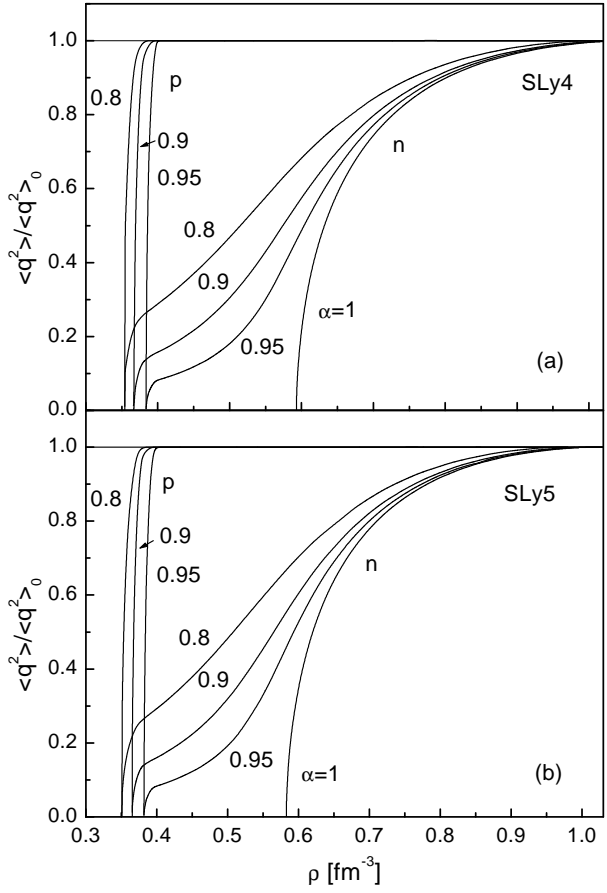} \caption{Same as in Fig.~\ref{fig1},
but for the second order moments $\langle {\bf q}^{2}\rangle_{p}$ and
$\langle {\bf q}^{2}\rangle_{n}$,  normalized to their values in the
totally polarized state. }\label{fig2}
\end{figure}

Another nontrivial feature relates to the density behavior of the
spin polarization parameters at large isospin asymmetry. As seen
from Fig.~1, even small admixture of protons leads to the
appearance of long tails in the density profiles of the neutron
spin polarization parameter near the transition point to a spin
ordered state. As a consequence, the spin polarized state is
formed much earlier in density than in pure neutron matter. For
example, the critical density in neutron matter is
$\varrho\approx0.59\,\mbox{fm}^{-3}$ for SLy4 potential and
$\varrho\approx0.58\,\mbox{fm}^{-3}$  for SLy5 potential; in
asymmetric nuclear matter with $\alpha=0.95$ the spin polarized
state arises at $\varrho\approx0.38\,\mbox{fm}^{-3}$ for SLy4
 and  SLy5 potentials. Hence, even small
quantity of protons strongly favors spin instability of highly
asymmetric nuclear matter, leading to the appearance of states
with the oppositely directed spins of  neutrons and protons. As
follows from Fig.~1, protons become totally spin polarized within
very narrow density domain (e.g., if $\alpha=0.95$, full
polarization occurs at $\varrho\approx0.41\,\mbox{fm}^{-3}$ for
SLy4 force and at $\varrho\approx0.40\,\mbox{fm}^{-3}$ for SLy5
force) while the threshold densities for the appearance and
saturation of the neutron spin order parameter are substantially
different (if $\alpha=0.95$, neutrons become totally polarized at
$\varrho\approx1.05\,\mbox{fm}^{-3}$ for SLy4 force and  at
$\varrho\approx1.02\,\mbox{fm}^{-3}$ for SLy5 force).

Note that the second order moments \bal \langle {\bf
q}^{2}\rangle_{n}&\equiv\langle {\bf
q}^{2}\rangle_{n\uparrow}-\langle {\bf
q}^{2}\rangle_{n\downarrow}\\
 &=\frac{1}{V}\sum_{\bf q} {\bf
q}^2\Bigl(n(\omega_{+,+})-n(\omega_{-,+})\Bigr),\nonumber\\
\langle {\bf q}^{2}\rangle_{p}&\equiv\langle {\bf
q}^{2}\rangle_{p\uparrow}-\langle {\bf
q}^{2}\rangle_{p\downarrow}\nonumber\\
 &=\frac{1}{V}\sum_{\bf q}
{\bf q}^2\Bigl(n(\omega_{+,-})-n(\omega_{-,-})\Bigr)
\nonumber\end{align}
\begin{figure}[t]
\includegraphics[height=12.6cm,width=8.6cm,trim=49mm 103mm 56mm 46mm,
draft=false,clip]{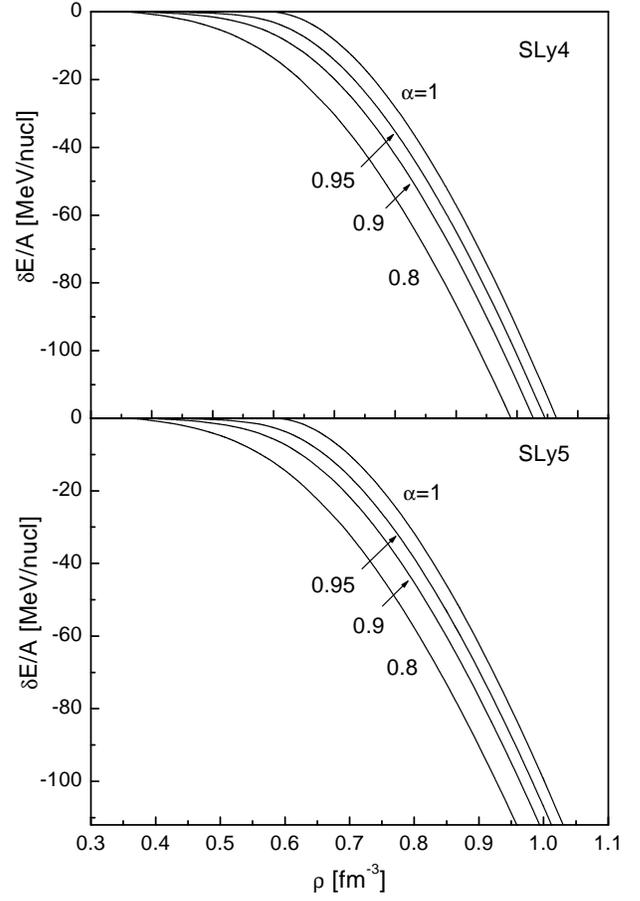} \caption{Total energy   per nucleon,
measured from its value in the normal state,  for the state with
the oppositely directed spins of neutrons and protons as a
function of density at zero temperature for (a) SLy4 force and (b)
SLy5 force. }\label{fig3}
\end{figure}
also characterize spin polarization of the neutron and proton
subsystems. If the solutions $\langle {\bf
q}^{2}\rangle_{30}$ and $\langle {\bf q}^{2}\rangle_{33}$ of the
self--consistent equations are known, then \bal \langle {\bf
q}^{2}\rangle_{n}&=\frac{1}{2}\bigl(\langle {\bf
q}^{2}\rangle_{30}+\langle {\bf q}^{2}\rangle_{33}\bigr),\nonumber\\
\langle {\bf q}^{2}\rangle_{p}&=\frac{1}{2}\bigl(\langle {\bf
q}^{2}\rangle_{30}-\langle {\bf
q}^{2}\rangle_{33}\bigr).\nonumber\end{align}

The values of $\langle {\bf q}^{2}\rangle_{n}$ and $\langle {\bf
q}^{2}\rangle_{p}$ for the totally polarized state are \beqestar
\langle {\bf q}^{2}\rangle_{n0}=\frac{3}{10}\varrho
k_F^2(1+\alpha)^{5/3},\;\langle {\bf
q}^{2}\rangle_{p0}=-\frac{3}{10}\varrho
k_F^2(1-\alpha)^{5/3}.\end{equation*}

In Fig.~\ref{fig2} we plot the density dependence of the second
order moments $\langle {\bf q}^{2}\rangle_{n}$ and $\langle {\bf
q}^{2}\rangle_{p}$, normalized to their values in the totally
polarized state, for different asymmetries at zero temperature.
These quantities behave similar to the spin polarization
parameters in Fig.~\ref{fig1}, i.e., there exist long tails in the
density profiles of the  neutron spin order parameter and the
proton spin order parameter is saturated within very narrow
density interval.

To check thermodynamic stability of the spin ordered state with
the oppositely directed spins of neutrons and protons, it is
necessary to compare the free energies of this state and the
normal state. In Fig.~\ref{fig3}  the difference of the total
energies per nucleon  of the spin ordered and normal states is
shown as a function of density at zero temperature. One can see
that nuclear matter undergoes a phase transition to the state with
the oppositely directed  spins of neutrons and protons at some
critical density, depending on the isospin asymmetry.

\section{Discussion and Conclusions}

Spin instability is a common feature, associated with a large
class of Skyrme models, but is not realized in more microscopic
calculations. The Skyrme interaction has been successful in
describing nuclei and their excited states. In addition, various
authors have exploited its applicability to describe bulk matter
at densities of relevance to neutron stars \cite{SMK}. The force
parameters are determined empirically by calculating the ground
state in the Hartree--Fock approximation and by fitting the
observed ground state properties of nuclei and nuclear matter at
the saturation density. In particular, the interaction parameters,
describing spin--spin and spin--isospin correlations, are
constrained from the data on isoscalar \cite{T,LS} and isovector
(giant Gamow--Teller)~\cite{SGE,BDE} spin--flip resonances.

In a microscopic approach, one starts with the bare interaction
and obtains an effective particle--hole interaction by solving
iteratively the Bethe--Goldstone equation. In contrast to the
Skyrme models, calculations with realistic NN potentials predict
more repulsive total energy per particle for a polarized state
comparing to the  nonpolarized one for all relevant densities,
and, hence, give no  indication of a phase transition to spin
ordered state. It must be emphasized that the interaction in the
spin-- and isospin--dependent channels is a crucial ingredient in
calculating spin properties of isospin asymmetric nuclear matter
and different behavior at high densities of the interaction
amplitudes, describing spin--spin and spin--isospin correlations,
lays behind this divergence in calculations with the effective and
realistic potentials.

In this study as a potential of NN interaction we choose SLy4 and
SLy5 Skyrme effective forces, which were constrained originally to
reproduce the results of microscopic neutron matter calculations
(pressure versus density curve)~\cite{CBH}. Besides, in the recent
publication~\cite{SMK} it was shown that the density dependence of
the nuclear symmetry energy, calculated up to densities
$\varrho\lesssim3\varrho_0$ with SLy4 and SLy5 effective forces
(together with some other sets of parameters among the total 87
Skyrme force parameterizations checked) gives the neutron star
models in a broad agreement with the observables, such as the
minimum rotation period, gravitational mass--radius relation, the
binding energy, released in supernova collapse, etc. This is
important check for using these parameterizations in high density
region of strongly asymmetric nuclear matter. However, it is
necessary to note, that the spin--dependent part of the Skyrme
interaction at densities of relevance to neutron stars still
remains  to be constrained. Probably, these constraints will be
obtained from the data on the time decay of magnetic field of
isolated neutron stars~\cite{PP}. In spite of this shortcoming,
SLy4 and SLy5 effective forces hold  one of the most competing
Skyrme parameterizations at present time for description of
isospin asymmetric nuclear matter at high density while a Fermi
liquid approach with Skyrme effective forces provides a consistent
and transparent framework for studying spin instabilities in a
nucleon system.

 In summary, we have considered the
possibility of phase transitions into spin ordered states of
strongly asymmetric nuclear matter within the Fermi liquid
formalism, where NN interaction is described by the Skyrme
effective forces (SLy4 and SLy5 potentials). In contrast to the
previous considerations, where the possibility of formation of FM
spin polarized states was studied on the base of calculation of
magnetic susceptibility,  we obtain the self--consistent equations
for the FM and AFM spin polarization parameters and solve them in
the case of zero temperature. It has been shown in the model  with
SLy4 and SLy5 effective forces, that strongly asymmetric nuclear
matter undergoes a phase transition to the spin polarized state
with the oppositely directed spins of neutrons and protons, while
the state with the same direction of the neutron and proton spins
does not appear. An important peculiarity of this phase transition
is the existence of long tails in the density profile of the
neutron spin polarization parameter near the transition point.
This means, that even small admixture of protons to neutron matter
leads to the considerable shift of the critical density of spin
instability in the direction of low densities. In the model with
SLy4 effective interaction this displacement is from the critical
density $\varrho\approx3.7\varrho_0$ for neutron matter to
$\varrho\approx2.4\varrho_0$ for asymmetric nuclear matter at the
isospin asymmetry $\alpha=0.95$, i.e. for $2.5\%$ of protons only.
As a result,  the state with the oppositely directed spins of
neutrons and protons appears, where protons become totally
polarized in a very narrow density domain. This picture is
different from the case of symmetric nuclear matter, where the FM
spin configuration is thermodynamically more preferable, than the
AFM one \cite{I}. Obtained results may be of importance for the
description of thermal and magnetic evolution of pulsars, whose
core represents strongly asymmetric nuclear matter.

A.I. is grateful for  support of Topical Program of APCTP during
his stay at Seoul. J.Y. is partially supported by Korea Research
Foundation Grant (KRF-2001-041-D00052).

\end{document}